\documentclass[showpacs,amssymb,superscriptaddress,notitlepage,nofootinbib,prd,longbibliography]{revtex4-2}
\usepackage{graphicx}
\usepackage{amsmath}
\usepackage{amsfonts}
\usepackage{bm}
\usepackage[colorlinks=true]{hyperref}
\usepackage[all]{hypcap}
\usepackage[utf8]{inputenc}
\usepackage{url}

\usepackage{float}

\usepackage{xcolor}
\usepackage{multirow}

\hypersetup{
    colorlinks,
    citecolor=[rgb]{0.16,0.53,0.80},
    linkcolor=[rgb]{0.16,0.03,0.50},
}

\input colordvi

\usepackage{tikz}

\usepackage{graphicx,color}

\def\be{\begin{equation}}
\def\ee{\end{equation}}
\def\bea{\begin{eqnarray}}
\def\eea{\end{eqnarray}}
\def\ptl{\partial}

\begin{document}

\centerline{\large{ \bf Vacuum polarization instead of "dark
matter" in a galaxy }}
\bigskip
\bigskip
\centerline{S.L. Cherkas $^a$, V.L. Kalashnikov$^b$}
\bigskip
\bigskip
\centerline{$^a$Institute for Nuclear Problems, Belarus State
University} \centerline{Minsk 220006, Belarus} \centerline{Email:
cherkas@inp.bsu.by}
\medskip
\centerline{$^b$ Department of Physics,}\centerline{  Norwegian
University of Science and
Technology,}\centerline{H{\o}gskoleringen 5, Realfagbygget, 7491,
Trondheim, Norway} \centerline{Email:
vladimir.kalashnikov@ntnu.no}
\bigskip
\centerline{Submitted: August 30, 2022}

\bigskip
\bigskip

\bigskip

{\fontsize{10}{11}\selectfont We considered a vacuum polarization
inside a galaxy in the eikonal approximation and found that two
possible types of polarization exist. The first type is described
by the equation of state $p=\rho/3$, similar to radiation. Using
the conformally-unimodular metric allows constructing a
nonsingular solution for this vacuum ``substance'', if a compact
astrophysical object exists in the galaxy's center. As a result,
a ``dark'' galactical halo appears that increases the rotation
velocity of a test particle as a function of the distance from a
galactic center. The second type of vacuum polarization has a more
complicated equation of state. As a static physical effect, it
produces renormalization of the gravitational constant, thus,
causing no static halo. However, a nonstationary polarization of
the second type, resulting from an exponential increase (or
decrease) of the galactic nuclei mass with time in some
hypothetical time-dependent process, produces a gravitational
potential looking like a dark matter halo.}

\bigskip

\section{Introduction}

Among the various issues of combining general relativity (GR) and
the quantum mechanics, one encounters the problems of vacuum
energy and black holes.

The first problem is to explain why enormous zero-point vacuum
energy density $\rho_v\sim k_{max}^4$ (here $k_{max}$ is the UV
energy scale of quantum field theory associated with a hard
3-momentum cutoff of the order of the Planck mass $M_p$) does not
influence a universe expansion (e.g., see
\cite{weinberg1989cosmological,peebles2003cosmological,most} and
references herein). The second problem is associated with the loss
of unitarity and information inside of the black hole horizon
(e.g., see \cite{Unruh2017,infor} and references therein), that
prevents the definition of a pure quantum state.

On the other hand, the basis of GR is a notion of manifold
\cite{mizn}, i.e., a metric space, which could be covered by
coordinate maps. When a concrete space-time possessing some
symmetry is considered, one aims to introduce a system of
coordinates allowing maximal covering of this particular manifold.
For instance, the Schwarzschild solution only describes the region
outside the horizon, and one has to introduce the Kruscal
coordinates to cover the complete domain \cite{lan}. Nevertheless,
one could admit an opposite view: restricting the manifold by
sewing all the black hole horizons by some coordinate
transformation. This approach is similar to a case when a man
finds a hole in his trousers at a knee. In such a case, he steps
back a little from the hole border and then subtends it into a
node with the help of sewing.

It is allowed using the conformally-unimodular (CUM) metric
\cite{ch1}, where
  an ultra-compact black hole-like astrophysical object appears as  a nonsingular ball named ``eicheons''
  \cite{Eicheons}. Besides, the vacuum energy problem could be partially solved
  in the CUM metric if one builds a gravity theory admitting an arbitrary
  choice of the energy density level \cite{ch1}. That is possible because the equations for
  evolution of the Hamiltonian $\mathcal H$ and the momentum constraints $\mathcal P$ admit
  not only the trivial solution  $\mathcal P = 0$, $\mathcal H = 0$,  but also  $\mathcal P = 0$, $\mathcal H =const$.
  The constant compensates for the main part of the vacuum energy density proportional to the Planck mass in the fourth degree  \cite{ch1,Haridasu}.
  Residual  energy density, remaining after omitting the main part of the vacuum energy density, is some kind of dark energy
  and results in a cosmological picture containing a period of linear evolution in cosmic time \cite{conf,Haridasu} followed by late
 accelerated expansion.

Both dark energy and dark matter (DM) are unknown ``substances''
appearing in  modern cosmology and astrophysics
\cite{iorio2006solar,Freese2009,Oks2021}. DM appears not only in
cosmological scales but also
 at  galaxy scales.
The lowest scale at which there is evidence for
 DM is of $\approx$ kpc \cite{weinberg2015cold,de2020dark}. Dark energy is associated with
 vacuum energy, whereas DM is expected to be some kind of a non-baryonic matter weakly
 interacting with the known particles of the standard model \cite{bertone2005particle,buchmueller2017search,2207.11330}. Nevertheless,  there
 are attempts to explain the DM by a DM-like behavior of vacuum energy \cite{albareti2014vacuum}, or a vacuum polarization induced by the gravitational field. Heuristic
 models of vacuum polarization like \cite{hajdukovic2012quantum,penner2016gravitational,Hajdukovic2019,fiscaletti2020dark,Penner2022}, which would
demand dipolar fluid \cite{blanchet2008model}, anti-gravitation
\cite{Chardin2021} or hydrodynamical phenomena in a vacuum treated
as hypothetical (super-)fluid
\cite{huang2016superfluid,sbitnev2015hydrodynamics,zloshchastiev2020alternative},
are of interest.

The conventional renormalization procedure of the quantum field
theory applied to vacuum energy near a massive object
\cite{hamber1995quantum,bonanno2000renormalization,ward2002quantum,kirilin2002quantum,Satz2005,Morley2021},
leads to modification of the gravitational potential only at small
distances of the order of gravitational radius that are
unobservable with current technologies. That is, the
renormalization excludes the manifestation of micro-scale
phenomena on the macro-scales (nevertheless, see
\cite{albareti2014vacuum}). This conclusion assumes the general
covariance of the mean vacuum value of stress-energy tensor
$<0|T_{\mu\nu}|0>$ on a curved background. However, the vacuum
state $|0>$, invariant relatively general transformation of
coordinates does not exist \cite{Birrell82}. That raises a
question: is it reasonable to demand the covariance of
$<0|T_{\mu\nu}|0>$ in the absence of invariant $|0>$? If
invariance violation, which implies the existence of ``\ae ther'',
takes place, then, like condensed-matter physics, DM still could
be treated as an emergent phenomenon produced by vacuum
polarization.

The outline of this paper is as follows. In Section \ref{uniform},
we argue the necessity of considering a vacuum polarization from a
cosmological point of view and explain that the CUM metric is
needed to omit the main part of vacuum energy. Section \ref{pert}
contains a perturbation formalism in the CUM metric, which is
required to introduce a vacuum polarization as some media, i.e.,
``\ae ther''. The eikonal approximation is used in Section
\ref{vacsec} to obtain the vacuum energy density and pressure of a
quantum scalar field by summating the contributions from the
distorted virtual plane waves. The expression for a vacuum
equation of state is obtained. In Section \ref{fpol}, the $F$-type
vacuum polarization, possessing a radiation equation of state, is
used in the Tolman-Volkov-Oppenheimer (TOV) equations for two
substances. This type of vacuum polarization results in a dark
halo if eicheon is situated in the galactic center. In Section
\ref{phipol}, the $\Phi$-type of vacuum polarization is
considered. This type of polarization leads to the renormalization
of the gravitational constant in the stationary case. But it can
contribute to the DM halo for the nonstationary processes. In
Conclusion, we summarize the consequences of two types of vacuum
polarization for galaxies. In Appendix, we consider the static and
empty universe to demonstrate an example of an exact solution for
the system of perturbations, taking into account the $F$-type
vacuum polarization.

\section{A spatially uniform universe in the CUM metric}
\label{uniform}

\subsection{CUM metric in the five vectors theory of gravity}

\noindent We based our analysis on using the CUM metric, which is
the foundation of the so-called five vectors theory (FVT)
\cite{ch1}.  In the course of this analysis, we will use the
particular cases of the CUM metric appropriate to the physics
considered.

A general class of CUM metrics  is defined as \cite{ch1}
\be
  ds^2\equiv g_{\mu\nu} dx^\mu dx^\nu = a^2\left(1-\ptl_m
P^m\right)^2d\eta^2-\gamma_{ij} (dx^i+ N^i d\eta) (dx^j+
N^jd\eta),
\label{interv1}
\ee
where $x^\mu=\{\eta,\bm x\}$, $\gamma_{ij}$ is a spatial metric,
$a =\gamma^{1/6}$ is a locally defined scale factor, and
$\gamma=\det\gamma_{ij}$, $\eta$ is a conformal time connected
with a cosmic time $t$ through $dt=a(\eta,\bm x) d\eta$.
 The spatial
part of the interval (\ref{interv1}) looks as
\be
dl^2\equiv\gamma_{ij}dx^idx^j=a^2(\eta,\bm x)\tilde
\gamma_{ij}dx^idx^j,
\label{interv2}
\ee
where  $\tilde\gamma_{ij}=\gamma_{ij}/a^2$ is a matrix with the
unit determinant. The interval (\ref {interv1}) is similar
formally to the ADM one \cite{adm}, but the lapse function is
taken in the form of $a(1-\ptl _ m P ^ m) $, where $P ^ m $ is a
three-dimensional vector, and $\ptl_m$ is a conventional partial
derivative. In the gravity theory admitting arbitrary choice of
the energy density level \cite{ch1}, there are the Lagrange
multipliers $\bm P$, the shift function $\bm N$, and three triads
$\bm e^{a}$ to parameterize the spatial metric
$\gamma_{ij}=e^a_ie^a_j$. Such model was named the FVT of gravity
\cite{ch1}. In contrast to GR, where the lapse and shift functions
are arbitrary, the restrictions $\ptl _ n(\ptl _ m N^m )=0$ and
$\ptl _ n(\ptl _ m P^m )=0$ arise in FVT. The Hamiltonian
$\mathcal H$ and momentum $\mathcal P_i$ constraints in  the
particular gauge $P^i=0$, $N^i=0$  obey the constraint evolution
equations \cite{ch1}:
\bea
\ptl_\eta{\mathcal H}=\ptl_i\left(\tilde \gamma^{ij}\mathcal P_j\right),\label{5} \\
\ptl_\eta {\mathcal P_i}=\frac{1}{3}\ptl_i {\mathcal H},\label{6}
\eea
which admits adding some constant to ${\mathcal H}$. In the FVT
frame, it is not necessary that $\mathcal H=0$, but  $\mathcal
H=const$ is also allowed. The particular cases of the CUM metric
corresponding to the Bianchi I model and the spherically symmetric
space-time were analyzed in \cite{bia,v2}.

\subsection{Uniform, isotropic and flat universe}

\noindent Let us consider a particular case of  (\ref{interv1})
\be
ds^2=a(\eta)^2(d\eta^2-d\bm x^2)
\ee
corresponding to a spatially uniform, isotropic and flat universe,
where
 the Friedmann equations take the form
\cite{conf,Cherkas07,eqofst1}:
\bea
M_p^{-2} e^{4 \alpha } \rho -\frac{1}{2} e^{2 \alpha}
\alpha ^{\prime 2}=const,\label{f1}\\
 \alpha ''+ \alpha^{\prime
2}=M_p^{-2} e^{2 \alpha} (\rho -3 p).\label{f2}
\eea
Here $\alpha(\eta)=\log a(\eta)$,  the prime denotes the
derivative in respect to the conformal time. We use the system of
units $\hbar=c=1$  and the reduced Planck mass
$M_p=\sqrt{\frac{3}{4\pi G}}$ (in physical units
$M_p=\sqrt{\frac{3\hbar c}{4\pi G}}$ ). According to FVT
\cite{ch1}, the first Friedmann equation (\ref{f1}) is satisfied
up to some constant, and the main parts of the vacuum energy
density and pressure
\bea
\rho_v\approx (N_{boson}-N_{ferm})\frac{k_{max}^4}{16\pi^2
a^4},\label{rhvac}
\\
p_v=\frac{1}{3}\rho_v
\label{pvac}
\eea
do not contribute to the universe expansion because the constant
in (\ref{f1})  compensates the vacuum energy density, whereas
there is no vacuum contribution in Eq. (\ref{f2}) by virtue of the
equation of state (\ref{pvac}).

Bosons and fermions contribute with opposite signs into a vacuum
energy density (\ref{rhvac}) \cite{Visser18,Visser2019}. Here, we
do not consider the supersymmetry hypotheses $N_{boson}-N_{ferm}$
due to the absence of evidence for supersymmetric partners to date
\cite{Workman:2022ynf}.

For the mass contributions of particles and condensates, we imply
the Pauli sum rules \cite{Visser2019,eqofst}. These rules are not
fulfilled at this moment due to the incompleteness of the standard
model. Nevertheless, one may hope they will be satisfied after
possible discoveries beyond the standard model.

Other contributors to the vacuum energy density are the terms
depending on the derivatives of the universe expansion rate
\cite{Cherkas07,eqofst,eqofst1,Haridasu}. They  have the correct
order of magnitude $\rho_{v}\sim M_p^2 H^2$, where $H$ is the
Hubble constant, and explain the accelerated expansion of the
universe driven by the residual energy density and pressure
\cite{Cherkas07,eqofst,eqofst1,Haridasu}:
\begin{equation}
\rho_v=\frac{a^{\prime 2}}{2a^6}M_p^2(2+N_{sc}){\mathcal
S}_0,~~~~p_v=\frac{M_p^2(2+N_{sc}){\mathcal
S}_0}{a^6}\left(\frac{1}{2}a^{\prime
2}-\frac{1}{3}a^{\prime\prime}a\right),
\label{eqn:rhoandp}
\end{equation}
where $
    {\mathcal S}_0 = \frac{k_{max}^2}{8 \pi^2 M_p^2}.$
    Eqs. (\ref{eqn:rhoandp}) include the number of minimally coupled scalar fields
$N_{sc}$ plus two degrees of freedom of the gravitational waves
\cite{Cherkas07}. The massless fermions and photons do not
contribute to (\ref{eqn:rhoandp}) \cite{Cherkas07}.

According to (\ref{eqn:rhoandp}),  the accelerated expansion of
universe
 allows finding a value of the momentum UV cut-off
\be
k_{max}\approx \frac{12 M_p}{\sqrt{2+N_{sc}}}
\label{kmax}
\ee
from the measured value of the universe deceleration parameter and
other cosmological observations \cite{Cherkas07,Haridasu}. It
should be noted that the UV cut-off of the 3-momentums $k_{max}$
in (\ref{rhvac}) and hereafter also reflects the diffeomorphism
symmetry violation \footnote{The CUM metric implies a preferred
time foliation of space-time. Using the CUM metric per se does not
predict some visible effects in the Solar system and all satellite
experiments if their results are expressed in a gauge invariant
way. At the same time, the use of the UV-cutoff at $k_{max}$
implies the Lorentz invariance violation. In the local particle
physics experiments, it leads to effects of the order of $\sim
\varepsilon/k_{max}\sim \varepsilon/M_{p}$, where $\varepsilon$ is
the typical energy of a particle, but certainly does not produce
some restrictions for Earth and satellite experiments. However, as
it will be shown below, consideration of vacuum physics using CUM
and $k_{max}$ could produce observable effects in a galaxy scale.}
(e.g., see
\cite{mattingly2005modern,amelino2013quantum,bluhm2021gravity,anber2010breaking,mavromatos2004cpt,h2}
and references herein).

\section{Perturbations of a uniform background in the CUM metric}
\label{pert}

\noindent In this section, the scalar perturbations\footnote{We
consider only scalar perturbations because the vector and tensor
perturbations do not perturb the matter.} of the CUM metric
(\ref{interv1}) are considered \cite{Cherkas2018}:
\be
  ds^2=a(\eta,\bm
x)^2\left(d\eta^2-\left(\left(1+\frac{1}{3}\sum_{m=1}^3
\ptl_m^2F(\eta,\bm x)\right)\delta_{ij}-\ptl_i\ptl_jF(\eta,\bm
x)\right)dx^idx^j\right).
\label{int1}
\ee
Here the perturbations of the locally defined scale factor are
expressed through a gravitational potential $\Phi$:
\be
a(\eta,\bm x)=e^{\alpha(\eta,\bm x)}\approx
e^{\alpha(\eta)}(1+\Phi(\eta,\bm x)).
\label{a}
\ee
 A
stress-energy tensor can be written in the hydrodynamic
approximation
\be
T_{\mu\nu}=(p+\rho)u_{\mu}u_{\nu}-p\, g_{\mu\nu}.
\label{tmn}
\ee
 The perturbations of the energy density $\rho(\eta,\bm
x)=\rho(\eta)+\delta \rho(\eta,\bm x)$ and pressure $p(\eta,\bm
x)=p(\eta)+\delta p(\eta,\bm x)$ are considered around spatially
uniform
 values.

 Let us  introduce new variables
\bea
\wp(\eta,\bm x)=a^4(\eta,\bm x)\rho(\eta,\bm x), \label{var1}\\
\Pi(\eta,\bm x)=a^4(\eta,\bm x)p(\eta,\bm x)\label{var2}
\eea
for  reasons which will be explained below. The perturbations of
(\ref{var1}), (\ref{var2}) around the uniform values can be
written now as $\wp(\eta,\bm
x)=e^{4\alpha(\eta)}\rho(\eta)+\delta\wp(\eta,\bm x)$,
$\Pi(\eta,\bm x)=e^{4\alpha(\eta)}p(\eta)+\delta\Pi(\eta,\bm x)$.
The 4-velocity $u$ is represented in the form of
\be
  u^{\mu}=e^{-\alpha(\eta)}\{1,\bm \nabla \frac{v(\eta,\bm x)}{\rho(\eta)
  +p(\eta)}\}\approx \{e^{-\alpha(\eta)}(1-\Phi(\eta,\bm
x)),e^{3\alpha(\eta)}\bm \nabla \frac{v(\eta,\bm
x)}{\wp(\eta)+\Pi(\eta)}\} ,
\label{vr}
\ee
where $v(\eta,\bm x)$ is a scalar function. Expanding all
perturbations into the Fourier series $\delta \wp(\eta,\bm
x)=\sum_{\bm k}\delta\wp_{\bm k}(\eta)e^{i \bm k \bm x }...$ etc.
results in the equations for the perturbations:

\bea
  -6  \Phi_{\bm k}'+6 \alpha' \Phi_{\bm k}+k^2  F_{\bm
k}'+\frac{18}{M_p^2}
 e^{-2\alpha}\sum_i v_{\bm k i}
=0,\label{con1}
\\
 -18 \alpha ' \Phi_{\bm k}^\prime -6 (k^2+3\alpha^{\prime 2})
\Phi_{\bm k} +k^4 F_{\bm k} +\frac{18}{M_p^2} e^{-2 \alpha }
\sum_i\delta\wp_{\bm k i}\,
=0,\label{con2}\\
 -12  \Phi_{\bm k}-3 ( F_{\bm k}''+2 \alpha'  F_{\bm k}')+k^2
 F_{\bm k}=0,\label{18}
\\
 -9 ( \Phi_{\bm k}''+2\alpha' \Phi_{\bm k}')-9(2\alpha''+2
\alpha '^2+ k^2) \Phi_{\bm k} +k^4
 F_{\bm k} -\frac{9}{M_p^2}e^{-2 \alpha
}\left(\sum_i 3 \delta \Pi_{\bm k i}-\delta  \wp_{\bm k
i}\right)=0,\label{eq4}
\eea
where the index $i$ denoting the kind of  substance has been
introduced. It is remarkable that, as a result of the choice of
the variables (\ref{var1}), (\ref{var2}), (\ref{vr}), the
unperturbed  values $\rho$ and $p$ do not appear in the system
(\ref{con1})-(\ref{eq4}). That allows us to avoid the influence of
the large uniform energy density and pressure (\ref{rhvac}),
(\ref{pvac}) on the evolution of perturbation. Eqs. (\ref{con1}),
(\ref{con2}) are consequences of the Hamiltonian and momentum
constraints, while  Eqs. (\ref{18}), (\ref{eq4}) are equations of
motion. For consistency of the constraints with the equations of
motion, every kind of fluid has to satisfy the continuity and
Euler  equations:
\bea
 \alpha'(\delta \wp_{\bm k i}-3 \delta \Pi_{\bm k
i})-(3\Pi_i-\wp_i)(\Phi_{\bm k}'-4\Phi_{\bm k}\alpha ')
+4\wp_i^\prime \Phi_{\bm k}-\delta \wp_{\bm k i}'+k^2  v_{\bm k i}=0,\label{eq5}\\
  \Phi_{\bm k}(\wp_i-3\Pi_i)+\delta  \Pi_{\bm k i}+ v_{\bm k
i}'=0.
\label{lasteq}
\eea

The equations (\ref{con1})-(\ref{lasteq}) have the same form as in
GR, but for the consistency of Hamiltonian and momentum
constraints (\ref{con1}), (\ref{con2}) with the equations of
motions (\ref{18})-(\ref{lasteq}), it is sufficient for the first
Friedmann equation (\ref{f1}) to be valid up to some constant.
Namely, for such consistency, it is necessary that the
differentiation of constraints with the subsequent substitution of
the second time derivatives from the equations of motion
(\ref{f2}), (\ref{18})-(\ref{lasteq}) lead to identical
equalities.
 This consistency is a feature
 of using the CUM metric, in particular, and the FVT theory, in general. In any other metrics
different from CUM (that is, in a frame of GR), the first
Friedmann equation (\ref{f1}) with the $const=0$ in the right hand
side is needed for consistency of the constraints and the
equations of motion.

\section{Vacuum as a medium: the eikonal approximation for quantum fields}
\label{vacsec}

\noindent Generally, a vacuum could also be considered as some
fluid (e.g., see
\cite{huang2016superfluid,sbitnev2015hydrodynamics,zloshchastiev2020alternative}),
i.e., ``\ae ther'' \cite{dirac1951there}, but with some stochastic
properties along with its elastic ones
\cite{eqofst,eqofst1,cherkas2021wave}. Here we are interested in
its elastic properties only. In Refs. \cite{eqofst,eqofst1}, the
speed of sound for the scalar  waves of vacuum polarization
$c_s^2=\frac{p_v^\prime(\eta)}{\rho_v^\prime(\eta)}$ was
introduced, where $p_v$ and $\rho_v$ are given by
(\ref{eqn:rhoandp}). That is only heuristic conjecture.

Here the actual calculations of the vacuum density and pressure on
the curved background are performed in
 the eikonal approximation. The last one has a very transparent background. In the Minkowski's
 space-time, the virtual plane waves penetrate space-time and, to obtain
 a vacuum energy density, we must summarize the contributions of
 every wave. In the curved space-time, it is necessary to
 summarize the contributions of the distorted waves to obtain
 the spatially non-uniform energy density and pressure.
 It should be mentioned that the eikonal approximation was successfully used in high
energy physics \cite{Czyz69} and even in gravity \cite{Kabat1992},
where the small-angle scattering amplitude of two massive
particles were calculated in all orders on gravitational constant
$G$.

A massless scalar field in the external gravitational field obeys
the equation

\be
\frac{1}{\sqrt{-g}}\ptl_\mu\left(\sqrt{-g}g^{\mu\nu}\ptl_\nu\right)\phi=0.
\label{1}
\ee
Using the gauge $\bm N=0$, $\bm P=0$ in (\ref{interv1}) reduces
the CUM metric to
\begin{equation}
ds^2 = a^2(d\eta ^2 - \tilde {\gamma }_{ij} dx^idx^j),
\label{eq31}
\ee
so that Eq. (\ref{1}) takes the form
\be
\phi^{\prime\prime}
+2\frac{a^\prime}{a}\phi^\prime-\frac{1}{a^2}\ptl_i\left(a^2\tilde\gamma^{ij}\ptl_j\right)\phi=0.
\label{26}
\ee
That leads to
\be
 \chi^{\prime\prime}-\chi\frac{a^{\prime\prime}}{a}-\tilde\gamma^{ij}\ptl_i\ptl_j\chi\
-\ptl_i\tilde\gamma^{ij}\ptl_j\,\chi+\frac{\chi}{a}\left(\tilde\gamma^{ij}\ptl_i\ptl_j
a+\ptl_ja\ptl_i\tilde\gamma^{ij}\right)=0
\label{eqchi}
\ee
after the change of variables $\phi=\chi/a$. Further, in the terms
of the metric perturbations $\Phi$ and $F$, we come to
\bea
\chi^{\prime\prime}-\Delta \chi+\hat V \chi=0,
\eea
where a ``potential'' operator $\hat V$ has the form
\bea
  \hat V=-\alpha^{\prime\prime}-\alpha^{\prime 2}-2
\alpha^\prime \Phi^\prime-
\Phi^{\prime\prime}+\Delta\Phi+\frac{1}{3}\Delta F\,\Delta-
\frac{\ptl^2F}{\ptl x^j\ptl x^i}\frac{\ptl^2}{\ptl x^j\ptl
x^i}-\frac{2}{3}(\bm \nabla(\Delta F))\cdot\bm \nabla~~.~~
\label{eqchi2}
\eea

A quantization of the scalar field in terms of creation and
annihilation operators implies \cite{Birrell82}
\be
\hat \chi(\eta,\bm x)=\sum_{\bm k}u_{\bm k}(\eta,\bm x)\hat
{\mbox{a}}_{\bm k}+u_{\bm k}^*(\eta,\bm x)\hat {\mbox{a}}_{\bm
k}^+,
\ee
where the function  $u_{\bm k}$ satisfies Eq.  (\ref{eqchi}), and
the orthogonality condition is \cite{Birrell82}
\be
\int (u_{\bm k}\ptl_\eta u_{\bm q}^* -u_{\bm k}^*\ptl_\eta u_{\bm
q} )d^3\bm x=i\delta_{\bm k\bm q}.
\ee
Solution of Eqs. (\ref{eqchi}), (\ref{eqchi2}) for the functions
$u_{\bm k}$ can be written in the eikonal approximation
\be
u_{\bm k}(\eta,\bm x)=\frac{1}{\sqrt{2 k}}\exp\left(-i\eta k+i\bm
k\bm x-i\Theta_{\bm k}(\eta,\bm x)\right),
\label{eik0}
\ee
which leads to the equation for the eikonal function
\bea
  2 k\Theta_{\bm k}^\prime+\left(2 k_m\tilde
\gamma^{mj}-i\ptl_m\tilde \gamma^{mj}\right)\ptl_j\Theta_{ \bm
k}+\frac{1}{a}\bigl(a^{\prime\prime}-\tilde\gamma^{ij}\ptl_i\ptl_j
a -\ptl_ja\ptl_i\tilde\gamma^{ij}\bigr)+i\,k_j\ptl_m \tilde h^{m
j}-k_m k_j\tilde h^{m j}=0 \label{uravn},
\eea
and, according to Eqs. (\ref{int1}), (\ref{a}), is written in the
terms of the metric perturbations $\Phi(\eta,\bm x)$, $F(\eta,\bm
x)$:
\be
 k \Theta_{\bm k}^\prime+\bm k\bm \nabla \Theta_{\bm
k}(\eta,\bm x) =\frac{1}{2} V_{\bm k},
\label{eikv}
\ee
where
\be
 V_{\bm k}(\eta,\bm x)=-2 \alpha^\prime \Phi^\prime -
\Phi^{\prime\prime}+\Delta\Phi+k_ik_j\ptl_i\ptl_jF-\frac{k^2}{3}\Delta
F.
\label{poten}
\ee
A solution of (\ref{eikv}) can be obtained in the form
\be
\Theta_{\bm k} (\eta,\bm x)=\frac{1}{2k}\int_{\eta_0}^\eta\,
V_{\bm k}\left(\tau,\bm x+\frac{\bm k}{k}(\tau-\eta)\right)d\tau,
\label{eikth}
\ee
where the lower integration limit $\eta_0$ depends on the
cosmological model. In particular, it could be $0$ or $-\infty$.
The mean value of the stress-energy tensor of a massless scalar
field
\be
\hat T_{\mu\nu}=\ptl_\mu\,{\hat \phi}\ptl_\nu{\hat
\phi}-\frac{1}{2}g_{\mu\nu}g^{\alpha\beta}\ptl_\alpha\,{\hat
\phi}\ptl_\beta\hat \phi
\label{tt}
\ee
can be averaged over the vacuum state and compared with the
hydrodynamic expression (\ref{tmn}). That gives
\bea
  \delta  \wp(\eta,\bm x)=e^{2\alpha(\eta)}<0|\frac{\hat
\phi^{\prime 2}}{2}+\frac{(\bm \nabla
\hat\phi)^2}{2}|0>\approx\frac{1}{2}\sum_{\bm k}
\frac{\alpha^\prime\Phi^\prime}{k}+\Theta_{\bm
k}^{\prime}-\frac{\bm k \bm \nabla \Theta_{\bm
k}}{k},\label{rho}\\
  \delta \Pi(\eta,\bm
x)=e^{2\alpha(\eta)}<0|\frac{\hat\phi^{\prime 2}}{2}-\frac{(\bm
\nabla \hat\phi)^2}{6}|0>\approx\frac{1}{2}\sum_{\bm
k}\frac{\alpha^\prime\Phi^\prime}{k}+ \Theta_{\bm
k}^\prime+\frac{\bm k\bm
\nabla\Theta_{\bm k}}{3 k},\label{pp}\\
 \bm \nabla v=-e^{2\alpha(\eta)}<0|\hat \phi^{\prime}\bm \nabla
{\hat \phi}|0>\approx\sum_{\bm k}\frac{\bm k\,\Theta^\prime_{\bm
k}}{k}-\bm\nabla\Theta_{\bm k}-\frac{\alpha^\prime\bm
\nabla\Phi}{k},\label{vv}
\eea
where  only spatially non-uniform parts of the vacuum averages are
implied in the second equalities on the right-hand side of
(\ref{rho}), (\ref{pp}) and (\ref{vv}). The last depends on the
metric perturbations $F(\eta,\bm x)$ and $\Phi(\eta,\bm x)$
contained in Eqs. (\ref{int1}), (\ref{a}). The final equalities in
(\ref{rho}), (\ref{pp}) and (\ref{vv}) result from calculations in
the eikonal approximation (\ref{eik0}).

Considering the quantity $\delta  \wp(\eta,\bm x)-3\delta
\Pi(\eta,\bm x)$ and using equations (\ref{eikv}) and
(\ref{poten}) result in
\bea
  \delta  \wp(\eta,\bm x)-3\delta  \Pi(\eta,\bm x)=-\sum_{\bm
k}\frac{\bm k \bm \nabla \Theta_{\bm k}}{k}+\Theta_{\bm
k}^{\prime}+\frac{\alpha^\prime\Phi^\prime}{k}=~~~~~~~~~~~~~~~~~~
~~~~~~~~~~~~~~~~~~~~~~~~~~~~~~~~~~~~~~\nonumber\\
  -\sum_{\bm k}\frac{1}{2k} V_{\bm
k}+\frac{\alpha^\prime\Phi^\prime}{k}=\sum_{\bm k}
\frac{1}{2k}\left(\Phi^{\prime\prime}-\Delta\Phi-k_ik_j\ptl_i\ptl_jF+\frac{k^2}{3}\Delta
F\right)=\frac{N_{sc}}{8\pi^2}k_{max}^2(\Phi^{\prime\prime}-\Delta\Phi),
\label{rho3p}
\eea
where summation has been changed by integration $\sum_{\bm
k}\rightarrow \int d^3\bm k/(2\pi)^3$ and it is taken into account
that
\par \noindent
$\int_{k<k_{max}}\frac{1}{2k}\left(k_ik_j-\frac{k^2}{3}\delta_{ij}\right)d^3\bm
k=0$. The  number $N_{sc}$ of the scalar fields minimally coupled
with gravity has been introduced also in (\ref{eqn:rhoandp}).

In consequence of Eq. (\ref{rho3p}), two types of spatially
nonuniform vacuum polarization exist. Namely, the $F$-polarization
has a radiation-type equation of state \footnote{For instance, see
a DM vacuum model with the equation of state ``running'' from
radiation-type to dark energy type \cite{albareti2014vacuum}.}
\be \delta \Pi_{vF}(\eta,\bm
x)=\frac{1}{3}\delta \wp_{vF}(\eta,\bm x),
\label{radst}
\ee
whereas the $\Phi$-polarization has an equation of state
\be \delta \Pi_{v\Phi}(\eta,\bm
x)=\frac{1}{3}\delta \wp_{v\Phi}(\eta,\bm
x)-\frac{N_{sc}}{24\pi^2}k_{max}^2(\Phi^{\prime\prime}-\Delta\Phi).
\label{phitype}
\ee
Both types of spatially nonuniform vacuum polarizations correspond
to the uniform component of (\ref{rhvac}), (\ref{pvac}), whereas
the uniform polarization given by (\ref{eqn:rhoandp}) has no
non-uniform counterpart with an accuracy of our consideration,
i.e., in the second-order on derivatives. It must be emphasized
that it is easy to obtain the equation of state (\ref{pvac}) for a
spatially
 uniform  main part of the vacuum energy density, but it is not so
trivial to do that  for a spatially non-uniform vacuum energy
density.

In principle, the system
(\ref{con1},\ref{con2},\ref{18},\ref{eq4},\ref{eq5},\ref{lasteq},\ref{radst},\ref{phitype})
is a fundamental system allowing to consider a broad range of
cosmological and astrophysical phenomena including CMB and BAO.
However below, we restrict ourselves to a galactic DM, which
scales from kpc to Mpc.

\section{Galactic DM as a $F$-vacuum polarization}
\label{fpol}

As it was shown in the section \ref{vacsec}, the $F$-component of
vacuum polarization has the equation of state analogous to
radiation (see Eq. \ref{pvac}). In this sense, it is similar to
the uniform part of vacuum energy density in Eq. (\ref{rhvac}).

At the same time, it is difficult to determine the concrete value
of the nonuniform vacuum energy density because, according to
(\ref{rho}), it contains an eikonal function $\Theta_{\bm k}$,
which is determined by the integral (\ref{eikth}). For instance,
one has $\Theta_{\bm k}(\eta,\bm r)=\sum_{\bm q} \tilde
\Theta_{\bm k,\bm q}(\eta)e^{i\bm q\bm r} $ from
(\ref{poten}),(\ref{eikth}) and

\be
 \tilde \Theta_{\bm k,\bm
q}(\eta)=\frac{1}{k}\left(\frac{1}{3}k^2q^2-(\bm q\bm
k)^2\right)\int_{\eta_0}^\eta F_{\bm q}(\tau)e^{i\bm k \bm
q(\tau-\eta)/k}d\tau.
\label{inthist}
\ee
Calculation of the integral (\ref{inthist}) requires one to know
the full evolution history of $F_{\bm q}(\tau)$. It is simpler to
use only the fact that the F-contribution to the vacuum
polarization has the  equation of state
\be
p_{vF}=\rho_{vF}/3.
\label{radf}
\ee The distributions of matter-energy density and potential are not determined for the static case in the first order on perturbations (see Appendix). However, it is possible to consider a
nonlinear heuristic model treating the $F$-vacuum as an abstract
substance with the above equation of state. The model consists of
a core of some incompressible substance, modeling a baryonic-like
matter placed on the radiation background, i.e., the $F$-polarized
vacuum or ``dark radiation'', which interacts with this core only
gravitationally. Below, we find a spherically symmetric solution
for an incompressible substance with the constant energy density
$\rho_1$ on the background of radiation density $\rho_2$.

\subsection{Equations in the CUM metric}
\noindent The CUM metric in the case of spherical symmetry
acquires the form \cite{Eicheons}
\begin{equation}
  ds^2 = a^2(d\eta ^2 - \tilde {\gamma }_{ij} dx^idx^j) =
e^{2\alpha }\left( {d\eta ^2 - e^{ - 2\lambda }(d{\bm x})^2 -
(e^{4\lambda } - e^{ - 2\lambda })({\bm x}d{\bm x})^2 / r^2}
\right),
\label{eq31a}
\ee
where  $r = \vert \bm x \vert $, $a = \exp \alpha $, and $\lambda
$ are  functions of $\eta , r$.  The matrix  $\tilde \gamma_{ij}$
with the unit determinant is expressed through $\lambda(\eta,r)$.
The interval (\ref{eq31a}) could be also rewritten in the
spherical coordinates:
\bea
x=r\sin \theta\cos\phi, ~~y=r\sin \theta\sin\phi,~~ z=r\cos\theta
\eea
to give
\begin{equation}
ds^2 = e^{2\alpha }\left( {d\eta ^2 - dr^2e^{4\lambda } - e^{ -
2\lambda }r^2\left( {d\theta ^2 + \sin^2\theta d\phi ^2} \right)}
\right){\kern 1pt} .
\label{eq42}
\end{equation}
Restricting ourselfves to static solutions, the equations for the
functions $\alpha(r)$ and $\lambda(r)$ are written as
\cite{Eicheons}
\bea
  \mathcal H=e^{2\alpha}\Biggl(   - \frac{e^{2\lambda }}{6r^2} +
e^{- 4\lambda }\Biggl( \frac{1}{6r^2} -
\frac{4}{3}\frac{d\alpha}{dr}\frac{d\lambda }{dr} +
\frac{1}{6}\left(\frac{d\alpha}{dr}\right) ^2 +
  \frac{2}{3r}\frac{d\alpha}{dr} +
\frac{1}{3}\frac{d^2\alpha}{dr^2} +~~~~~~~~~~~~~~~~\nonumber\\
\frac{7}{6}\left(\frac{d\lambda}{dr}\right)^2 -
\frac{5}{3r}\frac{d\lambda }{dr} -
\frac{1}{3}\frac{d^2\lambda}{dr^2}
\Biggr)+\frac{e^{2\alpha}}{{M_p^2}}\sum_j\rho_j(r)
   \Biggr) = const,
\label{eq32}
\eea
\bea
  \frac{d^2\alpha}{dr^2}= -\frac{3 e^{6 \lambda }}{r^2}+\frac{3}{r^2}-8 \frac{d\alpha}{dr} \frac{d\lambda}{dr}
 +7 \left(\frac{d\alpha}{dr}\right)^2+\frac{10}{r} \frac{d\alpha}{dr}+3 \left(\frac{d\lambda}{dr}\right)^2
 -\frac{6}{r} \frac{d\lambda}{dr}+3 \frac{e^{2 \alpha+4 \lambda}}{M_p^2}\sum_j\rho_j-3p_j,
\label{eq34}
 \eea
\bea
   \frac{d^2\lambda}{dr^2}= -\frac{5 e^{6 \lambda
}}{r^2}+\frac{5}{r^2}-18 \frac{d\alpha}{dr} \frac{d\lambda}{dr}+12
\left(\frac{d\alpha}{dr}\right)^2+\frac{18}{r}
\frac{d\alpha}{dr}+8
\left(\frac{d\lambda}{dr}\right)^2-\frac{14}{r}
\frac{d\lambda}{dr}+6 \frac{e^{2 \alpha+4
\lambda}}{M_p^2}\sum_j\rho_j-3p_j ,
\label{eq35}
\eea
where Eq. (\ref{eq32}) is the Hamiltonian constraint, which could
be rewritten in a form containing no second derivatives using Eqs.
(\ref{eq34}), (\ref{eq35}):
\bea
  \mathcal H=\frac{e^{2 \alpha-4 \lambda}}{2 r^2} {\left(-3 r^2
\left(\frac{d\alpha}{dr}\right)^2+4 r \frac{d\alpha}{dr} \left(r
\frac{d\lambda}{dr}-1\right)-\left(r
\frac{d\lambda}{dr}-1\right)^2+e^{6 \lambda
}\right)}+\frac{3e^{4\alpha}}{M_p^2}\sum_jp_j=const.
\label{h}
\eea
Each kind of substance has to satisfy
\be
\frac{d\,p_j}{dr}+(p_j+\rho_j)\frac{d\alpha}{dr}=0. \label{mat}
\ee

 A vacuum solution of the equations (\ref{eq32}), (\ref{eq34}), (\ref{eq35}) corresponding to the point massive particle was considered in \cite{Eicheons} where an absence of evidence for a
horizon was demonstrated. Let us consider another solution,
corresponding to the substance of a radiation type filling all the
space. This particular solution is written as
\be
\alpha(r)= \ln r-\frac{1}{6}\ln 7 ,~~~~~\lambda(r)=\frac{1}{6}{\ln
7},
\label{sol1}
\ee
and, under (\ref{radf}), it follows  from (\ref{mat}):
 \be
\frac{d}{dr}\left(\rho
e^{4\alpha}\right)=0,~~~~~~~~~\rho=\frac{1}{2}r^{-4}\,7^{-1/3},
\label{rhorad}
 \ee
if we use (\ref{sol1}) and (\ref{eq32}) with  $const=0$ in the
right hand side of Eq. (\ref{eq32}). Here, $\rho$ is measured in
the terms of $r_g^{-2} M_p^{-2}$, and $r$ is measured in the units
of $r_g$, which is not a gravitational radius, but some arbitrary
spatial scale. It should be noted that, for (\ref{radf}), Eqs.
(\ref{eq34}), (\ref{eq35}) look like those for an empty space,
whereas Eq. (\ref{eq32}) could also be considered as that for an
empty space, but with $const\ne0$. Thus, in the CUM metric of the
FVT where the Hamiltonian constraint is satisfied up to some
constant, one could alternatively consider the $F$-vacuum
polarization solution like that for an empty space, but with some
non-zero value of $const$ in Eqs. (\ref{eq32}), (\ref{h}).

Since the solution (\ref{rhorad}) is singular, it could be treated
as unphysical. To obtain a  realistic model,  one has to consider
at least two substances: a compact object in the center consisting
of a substance with a constant energy density and a substance with
the radiation equation of state (\ref{radst}). We must emphasize
the importance of such a dense kernel for obtaining non-singular
vacuum polarization of $F$-type.

\subsection{Equations in the Schwarzschild type metric}
\noindent It is more convenient to begin a consideration from the
Schwarzschild type metric \cite{wein}
\be
ds^2=B(R)dt^2-A(R)dR^2-R^2d\Omega,
\label{ab}
\ee
where Eqs. (\ref{sol1}), (\ref{rhorad}) correspond to the
well-known solution \cite{wein}
\be
\rho_2(R)=\frac{1}{14 R^2},
\label{solrad}
\ee
obeying the TOV equation \cite{tol,op} for a radiation fluid
\be
\rho_2'=-\frac{3 \rho_2 \left(m+4 \pi R^3 {\rho_2}/{3}\right)}{\pi
R \left(R-\frac{3 m}{2 \pi }\right)} \label{pp22eq}
\ee
 in all the spatial region $R\in(0,\infty)$,
where $m(R)$ is defined by
\be
m'(R)=4\pi  R^2 \rho_2.
\label{mout}
\ee

\begin{figure}[th]
\centering
  \includegraphics[width=14cm]{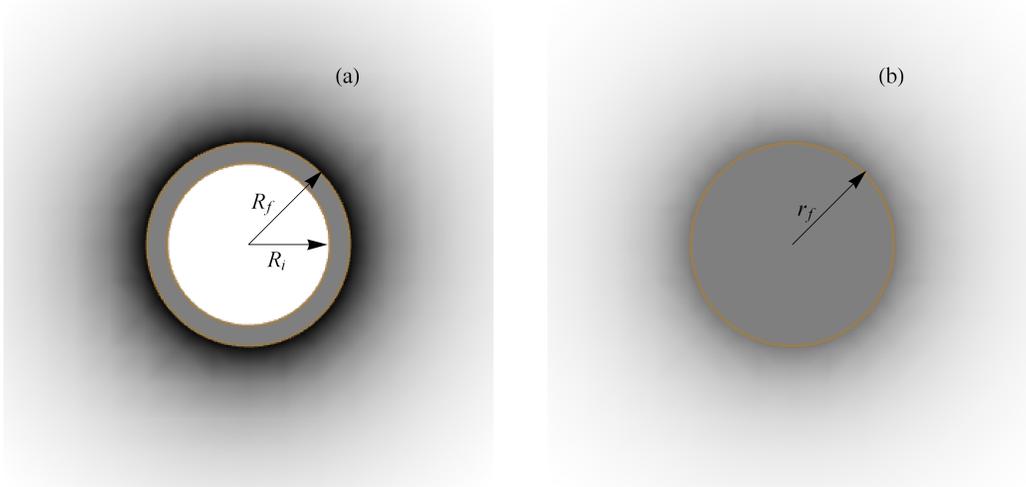}\\
  \caption{
(a) Schematic picture of an eicheon in the metric (\ref{ab})
taking into account a vacuum polarization in the form of dark
radiation,
  (b) an eicheon
in the metric (\ref{eq42}) looks like a solid sphere with a ``dark
radiation'' of the finite energy density in the center.
  }\label{piceich}
\end{figure}

\begin{figure}[th]
\centering
  \includegraphics[width=14cm]{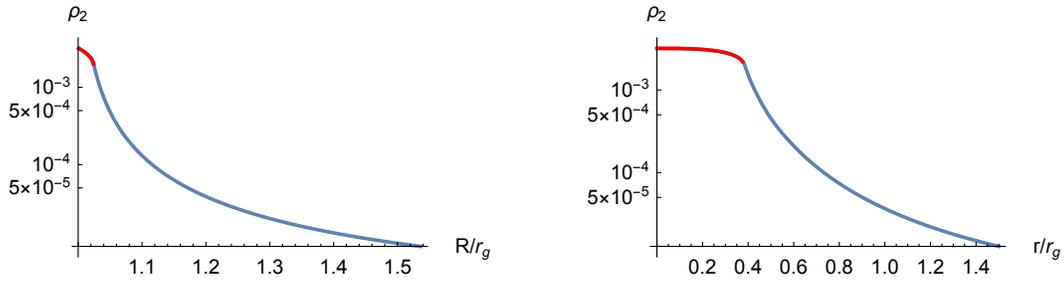}\\
  \caption{
(a) $\rho_2$ -- energy density of the vacuum polarization in a
form of ``dark radiation'' in the coordinates $R>R_i$ calculated
for the eicheon parameters $\rho_1=7 M_p^2r_g^{-2}$,
$R_i=1.001r_g$, $R_f=1.024r_g$, $\rho_2(R_f)=0.002 M_p^2r_g^{-2}$.
Red part of the curve corresponds to $R_i<R<R_f$, i.e., lies
inside an eicheon.  (b) $\rho_2$ calculated in
 the coordinates $r$ of the metric (\ref{eq42}). Red part of the curve corresponds to $0<r<r_f$.
  }\label{figrho}
\end{figure}

\begin{figure}[th]
\centering
  \includegraphics[width=8cm]{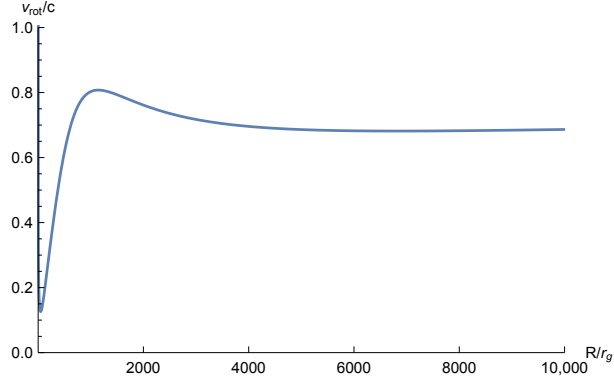}\\
  \caption{
   The general form of a model rotational curve for the eicheon parameters specified in the caption
   to Fig. \ref{figrho}.
  }\label{figtyp}
\end{figure}
\begin{figure}[th]
\centering
  \includegraphics[width=8cm]{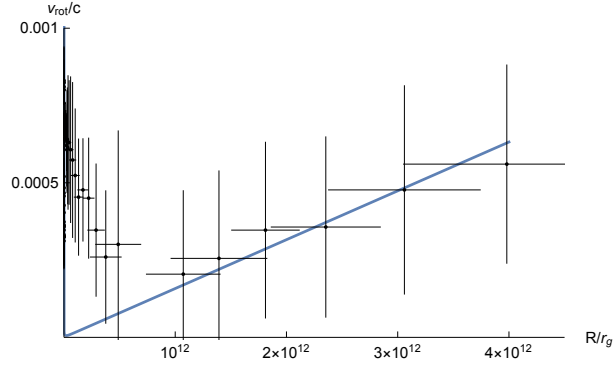}\\
  \caption{
   The rotational curve for the eicheon parameters $\rho_1=100 M_p^2r_g^{-2}$, $R_i=1.0001r_g$ and $\rho_2(R_f)=4\times 10^{-27}
   M_p^2r_g^{-2}$, where $r_g$ is defined by an eicheon mass.
   In the physical units $\rho_1=100 \frac{3c^6}{16\pi G^3m_\cdot^2}\approx 2.1\times 10^5~g/cm^3$.
   The points and error bars correspond to the Milky
Way rotational curve from \cite{Sofue}.
  }\label{vrot0}
\end{figure}
\begin{figure}[th]
\centering
  \includegraphics[width=9cm]{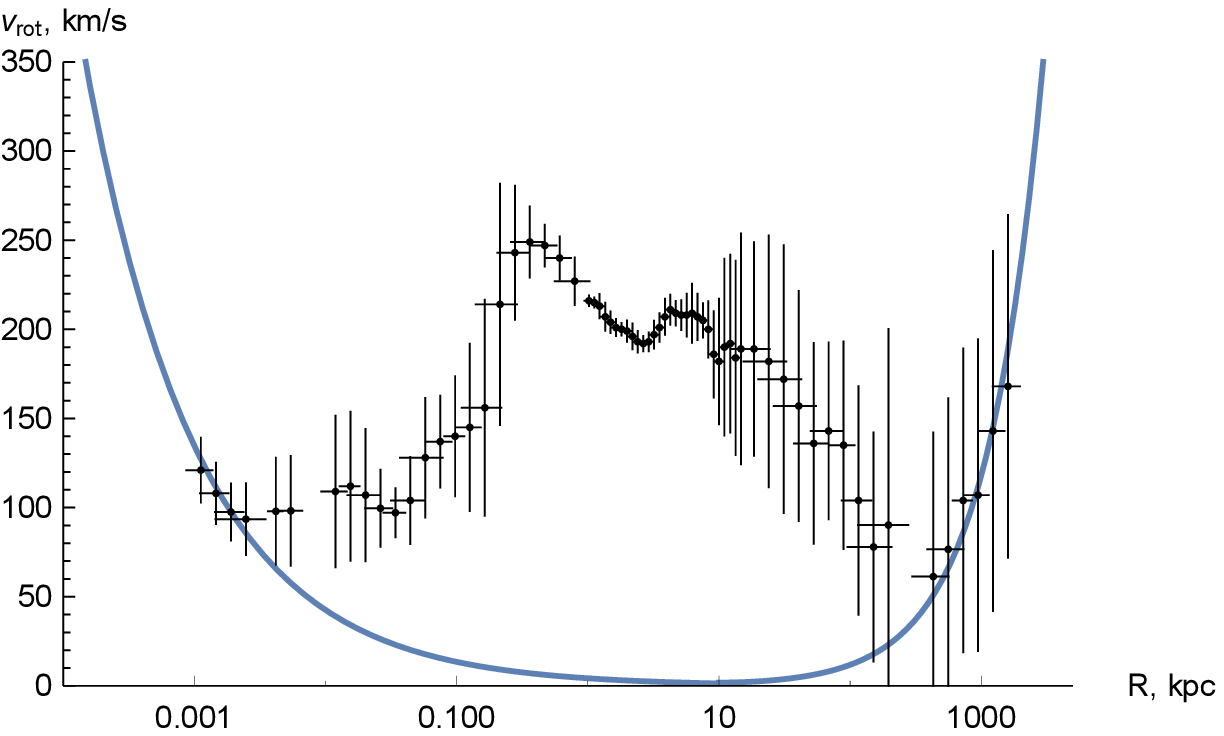}\\
  \caption{
The rotational curve of eicheon with the mass of Sgr A$^*$ with
taking into account the vacuum polarization of $F$-type.   The
logarithmic scale is used and the points correspond to the Milky
Way rotational curve from \cite{Sofue}. The eicheon  parameters
are given in the caption to Fig. \ref{vrot0}.
  }\label{mwdm}
\end{figure}
\noindent Again, $\rho_2$ is measured in the terms of
$r_g^{-2}M_p^2$, and $R$ is measured in the units of $r_g$. The
solutions (\ref{solrad}) and (\ref{rhorad}) are singular at $R=0$
and, thereby, nonphysical.  The situation changes cardinally in
the presence of a core consisting of incompressible matter. More
exactly, in the presence of incompressible matter of low density
$\rho_1$, the corresponding solution remains singular. However, if
$\rho_1>\frac{1}{2}\left(\frac{8}{9}\right)$, a solid ball in the
metric (\ref{eq42}) looks like a shell over $r_g$ in the metric
(\ref{ab}) \cite{Eicheons} that is shown in Fig. \ref{piceich}
(a).  Here, we again imply the gravitational radius $r_g$ as a
measure of the distances, but calculate it taking into account
only an incompressible matter. Such a matter occupies a region
between $R_i$ and $R_f$, where
\be
R_f=\sqrt[3]{R_i^3+\frac{1}{2\rho_1}}
\label{finpoi}
\ee in the units of $r_g$. Here the energy density $\rho_1$ is
constant and measured in the terms of $r_g^{-2}M_p^2$, where the
gravitational radius is defined as $r_g=\frac{3m_1}{2\pi M_p^2}$
and $m_1=\frac{4}{3}\pi\rho_1(R_f^3-R_i^3)$. Compact object of
such a type arising in FVT is known as "eicheon" \cite{Eicheons}
and replaces a black hole of GR. The appearance of eicheon in the
center makes the solution (\ref{pp22eq}) to be nonsingular because
it allows for setting the finite boundary conditions for
radiation.

To explain this, let us  consider two fluids in the metric
(\ref{ab}) obeying the TOV equations:
\bea
p_1'=-\frac{3(p_1+\rho_1) \left(m+4 \pi R^3
\left(p_1+\frac{\rho_2}{3}\right)\right)}{4 \pi  R \left(R-\frac{3
m}{2 \pi  }\right)},\label{p1eq}\\
\rho_2'=-\frac{3 \rho_2 \left(m+4 \pi R^3
\left(p_1+\frac{\rho_2}{3}\right)\right)}{\pi R \left(R-\frac{3
m}{2 \pi  }\right)},\label{p2eq}
\eea
where the function $m(R)$ satisfies to
\be
m'(R)=4\pi  R^2 (\rho_1+\rho_2).
\label{minn}
\ee

\noindent For  $\rho_1>\frac{1}{2}\left(\frac{8}{9}\right)$, the
above equations hold for the internal range $R_i<R<R_f$, where
$R_i>r_g$ and the border, occupied by $\rho_1$, is defined through
(\ref{finpoi}).

The pressure of incompressible fluid must turn to zero at the edge
of the range filled by matter $R=R_f$, and it is a boundary
condition for $p_1$. Then, one could set an amount of radiation at
$R=R_f$ and solve the system of equations in a region of
$\{R_i,R_f\}$ assuming $m(R_i)=0$. A solution allows determining
$m(R_f)$, and, using this value as an initial condition, one
should solve the equation for the radiation fluid (\ref{pp22eq})
in an outer region of $\{R_f,\infty\}$. The metric obtained by
solving the equations is \cite{wein}
\bea
\frac{1}{B}\frac{dB}{dR}=-\frac{2}{p_1+\rho_1}\frac{dp_1}{dR}=-\frac{2}{p_2+\rho_2}\frac{dp_2}{dR},\\
\frac{d}{dR}\left(\frac{R}{A}\right)=1-{6}\,R^2(\rho_1+\rho_2).
\eea
Comparing the   metric (\ref{eq31a}) and (\ref{ab}) leads to
relation for the radial coordinates $R$ and $r$ \cite{Eicheons}
\be
\frac{dR}{dr}=\left(\frac{r}{R}\right)^2\frac{B^{3/2}}{A^{1/2}},
\label{depr}
\ee
where the dependencies $B(R(r))$ and $A(R(r))$ are implied. Eq.
(\ref{depr}) has to be integrated with the initial condition
$R(0)=R_i$, which means that $R_i$ in the metric (\ref{ab})
corresponds to $r=0$ in the metric (\ref{eq31a}). Knowing $R(r)$
allows plotting $\rho_2(R)$ shown in Fig. 2 (a) as the $r$--
dependent function $\rho_2(R(r))$ (Fig. 2 (b)).

Let us consider the motion of a test particle on a circular orbit
in the metric (\ref{ab}). The angular velocity on a circular orbit
is calculated as \cite{wein}:
\be
\frac{d\phi}{dt}=\sqrt{\frac{1}{2R}\frac{dB}{dR}}.
\ee
A spatial interval followed by a particle along the circular orbit
is given by $ dl=R d\phi=R\frac{d\phi}{dt}dt $. To obtain the
rotation velocity observed by an observer situated at rest near
the moving particle, one has to divide the spatial interval over
the proper time $\sqrt{g_{00}}dt=\sqrt{B}dt$ of such an observer
\cite{rah}:
\be
v_{rot}=\frac{dl}{\sqrt{B}dt}=\sqrt{\frac{R}{2B}\frac{dB}{dR}}=
\sqrt{-\frac{R}{p_2+\rho_2}\frac{dp_2}{dR}}=\frac{1}{2}\sqrt{-\frac{R}{\rho_2}\frac{d\rho_2}{dR}}.
\ee

A qualitative example of the general form of the numerical
solution for the rotation velocity is shown in Fig. \ref{figtyp}.
Although the shape of the curve resembles observational data,
asymptotic of the rotation curve corresponds to $v_{rot}\sim
1/\sqrt{2}\approx 0.71$. This very large velocity (in units
 of speed of light) corresponds to asymptotic value $\rho_2\sim R^{-2}$ in (\ref{solrad}), whereas,
 in the reality, the rotation velocities of galaxies are $v_{rot}\sim 100-300~$km/s$~\sim 0.001$.
 To obtain smaller velocities, one has to diminish density of radiation in the center of eicheon, i.e.
 at $r=0$ in the metric (\ref{eq42}) or $R=R_i$ in the metric (\ref{ab}). For central radiation density of $\rho_2=4.6\times
10^{-27}~M_p^2r_g^{-2}=9.6\times10^{-24}~$g/$\mbox{cm}^3$, one has
the rotation curve shown in Fig. \ref{vrot0}. That is a pure
``dark radiation'' contribution without the galaxy bulge or disk.
It increases linearly with the distance and corresponds to the
rising part of the general curve shown in Fig. \ref{figtyp}. In
the logarithmic scale, one could see (Fig. \ref{mwdm}) together
the contribution of the eicheon of the mass of $4.2\times
10^6~M_{\bigodot}$ in the center of the Milky Way (the left side
of the curve) and the impact of the dark radiation (the right side
of the curve), whereas the effects of the galactic bulge and disk
responsible for the intermediate region are not taken into
account.   However, it is expected that bulge and disk attraction
will influence the $F$-type vacuum polarization in such a way that
the curve in Fig. \ref{vrot0} will be not pure linear but slightly
bent. We do not gain insight into such details because our goal is
to show that the $F$-type vacuum polarization could arise only
around a ``sewed'' black hole, i.e., around eicheon.

We emphasize that the presented consideration is heuristic
because, although the linear system for the perturbation and the
eikonal approximation for vacuum polarization seems reasonable, we
use its results in the nonlinear TOV model. Another thing is that
we set the density of radiation (the $F$-type vacuum polarization)
in the center of eicheon, i.e., at $r=0$, of $R=R_i$ empirically
but not calculate it from the first principles, i.e., we use only
the equation of state obtained from the  calculations in the
eikonal approximation.

\section{Vacuum polarization of $\Phi$-type}
\label{phipol}

In Sections \ref{pert}, \ref{vacsec} the linear system of equation
(\ref{18}),(\ref{eq4}),(\ref{eq5}),(\ref{lasteq}),(\ref{rho3p})
was deduced, which describes the evolution of perturbation by
taking into account vacuum polarization (see Eq. (\ref{rho3p})).
Galaxy formation is a complex nonlinear process that develops over
cosmological time scales. Generally, the linear system is
insufficient to describe the galaxy evolution. However, one could
create a heuristic picture setting an approximate profile of
matter near the galaxy center and obtain a gravitational potential
produced by vacuum polarization obeying the linear equations.
Below we will discuss that the observed galaxy halo could
originate from a very fast  (compared to the cosmological times)
growth of the galactic nucleus mass. We will neglect a
cosmological evolution assuming $\alpha(\eta)=0$. That reduces the
above system of the equations to
\bea
 -12 \Phi_{\bm q}-3  F_{\bm q}''+q^2  F_{\bm q}=0,\label{18al}
\\
 -9 \Phi_{\bm q}''-9 q^2  \Phi_{\bm q} +q^4  F_{\bm q} +
\frac{9}{M_p^2}\left(\sum_i \delta  \wp_{\bm k i}-3 \delta
\Pi_{\bm q i}\right)=0.\label{eq4al}\\
  \delta  \wp_{\bm q v}-3 \delta \Pi_{\bm q
v}=\frac{N_{sc}}{8\pi^2}k_{max}^2(\Phi^{\prime\prime}_{\bm
q}+q^2\Phi_{\bm q}),
\label{vac}
\eea
where the last equation holds  for the vacuum polarization of
$\Phi$--type and is denoted by $i=v$. Substituting $\Phi_{\bm q}$
from Eq. (\ref{18al}), and $ \delta \wp_{\bm q v}-3 \delta
\Pi_{\bm q v}$ from Eq. (\ref{vac}) into Eq. (\ref{eq4al}) gives
the equation
\be
  3 \left({k_{max}}^2-8 \pi ^2 {M_p}^2\right) \left(3
F^{\prime\prime\prime\prime}_{\bm q}+2 q^2 F_{\bm
q}^{\prime\prime}\right)-q^4 F_{\bm q} \left(3 N_{sc}{k_{max}}^2+8
\pi ^2 {M_p}^2\right)=288 \pi ^2{\delta \wp_{\bm q \,eff}}(\eta ),
\label{eff}
\ee
where an effective ``density'' of all the substances except vacuum
is denoted as
\be
\delta \wp_{\bm q\, eff}(\eta )=\sum_{i\ne v} \delta  \wp_{\bm k
i}-3 \delta \Pi_{\bm q i}.
\ee
Eq. (\ref{eff}) allows developing a simple model: setting profile
and time dependencies of the quantity $\wp_{\bm q\, eff}(\eta)$
empirically allows finding the metric perturbation $F_{\bm q}$ and
calculate $\Phi_{\bm q}$ using (\ref{18al}), i.e., to determine
the gravitational field corresponding to $\wp_{\bm q\,
eff}(\eta)$.

Let us, for simplicity, take $\wp_{\bm q\, eff}(\eta)$ in the form
of
\be
 \wp_{\bm q\,
eff}(\eta)=m\, Z(q)e^{\eta/T},
\label{empir}
\ee
where $m$ is a ``mass'' of the object at $\eta=0$, $Z(q)$ is a
form-factor and $T$ is some typical time of the ``mass'' growth.
 The  model implies   some rapid processes like
accretion occurring around the massive object, i.e., around the
galaxy nucleus. Substitution of the expression (\ref{empir}) into
Eq. (\ref{eff}) allows finding $F_{\bm q}(\eta)=\tilde F_{\bm
q}e^{\eta/T}$, where
\be
\tilde F_{\bm q}=-\frac{288 \pi ^2 T^4 m Z(q)}{3N_{sc} {k_{max}}^2
\left(q^4 T^4-2 q^2 T^2-3\right)+8 \pi ^2 {M_p}^2 \left(q^2
T^2+3\right)^2},
\label{ffq}
\ee
and Eq. (\ref{18al}) give $\Phi_{\bm q}(\eta)=\tilde \Phi_{\bm
q}e^{\eta/T}$:
\be
\tilde \Phi_{\bm q}=-\frac{24 \pi ^2 T^2 \left(q^2 T^2-3\right) m
Z(q)}{3N_{sc} {k_{max}}^2 \left(q^4 T^4-2 q^2 T^2-3\right)+8 \pi
^2 {M_p}^2 \left(q^2 T^2+3\right)^2}. \label{ppq}
\ee
At  $T\rightarrow\infty$, the corresponding  static limit is
\be
\tilde \Phi_{\bm q}=-\frac{24 \pi ^2 m Z(q)}{(3N_{sc}
{k_{max}}^2+8 \pi ^2 {M_p}^2) q^2}, \label{ppq1}
\ee
which implies that the vacuum polarization leads to
renormalization (increasing) of the Planck mass, i.e., decreasing
the gravitational constant. In particular, using the value
(\ref{kmax}) obtained from the cosmological observations
\cite{Haridasu} gives
\be
M_{p\,ren}^2=\left(1+\frac{54N_{sc}}{\pi^2(2+N_{sc})}\right)M_p^2,~~~~~~~~~G_{ren}=G/\left(1+\frac{54N_{sc}}{\pi^2(2+N_{sc})}\right).
\ee
It seems that the vacuum polarization, in some sense, acts like
antigravitation, and the gravitational constant $G_{ren}$
appearing in Newton's law has to differ from the gravitational
constant $G$ in the Friedmann equations for a uniform universe.
Although the gravitational constant's renormalization does not
influence the cosmological balance of the different kinds of
matter expressed in the units of the critical density $M_p^2H^2$,
it should be taken into account for comparison with the directly
measured (for instance, utilizing luminosity) density. Numerically
$N_{sc}=2$ gives $G_{ren}\approx 0.27\, G$.

\subsection{Invariant potentials and rotational curves}

Astrophysicists  express the results of observations in terms of
gauge-invariant quantities, which are not dependent on system of
coordinates. The potentials $\Phi(\eta,\bm x)$ and $F(\eta,\bm x)$
are not invariant relatively the infinitesimal transformations of
coordinates and time of the following type
\be
t=\eta+\xi_1(\eta,\bm x),~~~~~~~~~~~~~ \bm r=\bm x+\bm \nabla
\xi_2(\eta, \bm x),
\label{trans}
\ee
where $\xi_1(\eta,\bm x)$ and $\xi_2(\eta,\bm x)$ are some small
functions. Usually the potentials $\Phi_{inv}(\eta,\bm x)$ and
$\Psi_{inv}(\eta,\bm x)$ are introduced \cite{rio,hu,mukh} which
are invariant relatively transformations (\ref{trans}). The
potentials correspond to the metric
\be
ds^2=a^2(\eta)\left((1+2\Phi_{inv}(\eta,\bm
x))d\eta^2-\left(1-2\Psi_{inv}(\eta,\bm
x)\right)\delta_{ij}dx^idx^j\right)
\label{muhmet}
\ee
and are expressed through $\Phi$ and $F$ as
\bea
  \Phi_{\bm q\,inv}(\eta)=\Phi_{\bm q}(\eta)+\frac{a'(\eta ) F_{\bm
q}'(\eta
)+a(\eta ) F_{\bm q}''(\eta )}{2 a(\eta )}=\Phi_{\bm q}+\frac{F_{\bm q}}{2T^2}, \label{invq1}\\
  \Psi_{\bm q\,inv}(\eta)=-\frac{a'(\eta ) F_{\bm q}'(\eta )}{2
a(\eta )}-\Phi_{\bm q}(\eta )+\frac{1}{6} q^2 F_{\bm
q}(\eta)=-\Phi_{\bm q}(\eta )+\frac{1}{6} q^2 F_{\bm q},
\label{invq2}
\eea
where the final equalities at the right-hand side of
(\ref{invq1}), (\ref{invq2})  hold for our case of $a=const$, and
$\Phi,F\sim \exp\left(\eta/T\right)$. Using (\ref{ffq}),
(\ref{ppq}) gives
\be
\tilde \Phi_{\bm q\,inv}=-\frac{24 \pi ^2 T^2 \left(q^2
T^2+3\right) m \,Z(q)}{3N_{sc}\, {k_{max}}^2 \left(q^4 T^4-2 q^2
T^2-3\right)+8 \pi ^2 {M_p}^2 \left(q^2 T^2+3\right)^2},
\ee
and $\tilde \Psi_{\bm q\,inv}=\tilde \Phi_{\bm q\,inv}$. Thus, we
obtained the Fourier transformation of the time-dependent
gravitational potential $ \Phi_{\bm q\,inv}=\tilde \Phi_{\bm
q\,inv}e^{\eta/T}$ allowing us to define
\be
\Phi_{inv}(\bm x,\eta)=\frac{e^{\eta/T}}{(2\pi)^3}\int \tilde
\Phi_{\bm q\,inv}\,e^{i\bm q\bm x} d^3\bm q.
\label{inv}
\ee
To obtain a concrete empirical formula, one has to set the form
factor $Z(q)$, for instance, using the Gaussian profile $ \delta
\tilde \wp_{eff}(\bm x)=\pi^{-3/2}m\, D^{-3}\, e^{-x^2/D^2}$. The
spatial dependence of the potential (\ref{inv}) at the present
time, i.e., $\eta=0$ allows us to find the rotational velocity
dependence on the spatial coordinate
\be
v_{rot}(r)=\sqrt{-r\frac{d\Phi_{inv}(r)}{dr}}.
\ee
Here, the potential (\ref{inv}) is time-dependent, and actually,
there are no pure rotational curves because the radial velocities
are present. Here, for an estimation,  we discuss only tangential
velocity. The parameters $m$, $D$ and $Z(q)$ are adopted to
produce a typical rotational curve without an DM (blue curve in
Fig. \ref{fig6}), then vacuum polarization produces a halo
corresponding to black curve in Fig.  \ref{fig6}.
\begin{figure}[th]
 \begin{center}
  \includegraphics[width=9cm]{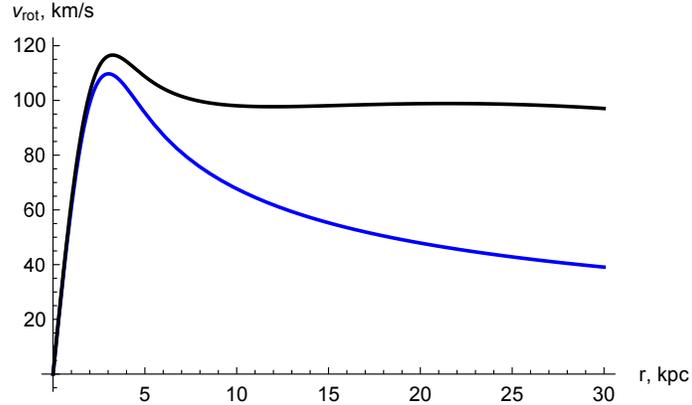}
  \end{center}
  \caption{
     The lower blue curve corresponds to the contribution of a galactic
  nucleus of baryonic matter, including specifically nucleus, bulge, and disk. The upper black curve takes the vacuum $\Phi$-polarization into account. The form factor of a galaxy nuclei is taken as $Z(q)-\exp\left(-\lambda q^2\right)$, $\lambda=1$,  the accretion rate is of $T=10$, i.e., $10$ kPc, which corresponds to 32000  years. Number of the minimally coupled scalar fields is of
   $N_{sc}=2$, and $k_{max}^2=8M_p^2\pi^2\frac{98}{100}$ is assumed.}
  \label{fig6}.
\end{figure}
The rotational curve has some similarities with the conventional
picture at $N_{sc}=2$, but in the conventional picture, the
contribution of the galactic nucleus, bulge, and disk are taken
into account. We include all these components into the Gaussian
form factor of galactic baryonic skeleton and call it "nucleus" in
our oversimplified picture. Then we permit it to increase (or
decrease) with time and obtain vacuum polarization caused by this
process.

\section{Conclusion}
We have considered two types of vacuum polarization corresponding
to the $F$- and $\Phi$-types of metric perturbations in the CUM
frame.

The $F$-type of spatially-nonuniform vacuum polarization
 has the radiation-type equation of state. In the first order
 on perturbations, it is impossible to determine a form of the static gravitational potential around an astrophysical object. In the frameworks of a nonlinear heuristic model using the TOV equations for matter and radiation, it was found that the solution, which is nonsingular at $r=0$, only arises if an eicheon is present in the galaxy's center. Eicheon is an analog of the black hole in GR and looks like an empty ``nut'' in the Schwarzschild type metric. From this point of view, we assume that DM, as a vacuum polarization, arises only in the galaxies having an eicheon (i.e., a ``black hole-like'' object) in the center. Namely, the eicheon conjecture allows us to convert a singular solution for pure radiation into a nonsingular physical one. Galaxies without an eicheon in
 the center (e.g., diffuse galaxies) is not to have a DM halo \footnote{
 The diffuse galaxy NGC1052-DF2
 \cite{VanDokkum} seems to contain no DM, whereas another
  diffuse galaxy Dragonfly 44  is supposed to contain a lot of DM \cite{van2016high}. However, for the last, we do know definitely, whether or not there is an eicheon in its center. }.

Under the oversimplified assumption of an isolated galaxy, the
dark halo, in terms of a test particle's rotation velocity, always
increases with the distance from the galaxy's center. Decreasing
the halo could occur only due to a violation of the galaxy's
isolation, i.e., at the distance of $\sim ~$2 Mpc. It should be
noted that the Andromeda galaxy is only 0.7 Mpc away. Generally,
the galaxies tend to form clusters. These evident facts urge the
development of a model of interacting galaxies with vacuum
polarization.

 For the $\Phi$-type of vacuum polarization, the renormalization of the gravitational constant (or Planck mass) has been found. That
 means that the gravitational constant found in the Earth, the Solar system, and galaxy observations is not equal (approximately four times less) to the gravitational constant used in cosmology to describe a spatially uniform universe. This fact does not influence the balance of the different kinds of matter in cosmology if one measures them in $M_p^2H^2$. Nevertheless, it increases the directly counted matter contribution fourfold, i.e., the luminous baryonic matter has to contribute $3.7$-times stronger into the cosmological Friedmann equations.

 The second effect of the $\Phi$-type polarization is the creation of the dark halo in the nonstationary process. It is found that the time-dependent evolving mass of the galaxy nuclei produces the gravitational potential of the dark halo type. This point urges a more careful observational investigation of the possible nonstationary origin of the dark halo. However, the required time for the galaxy nuclei mass growth seems very short $\sim 32000$ years. In such a situation, clarifying the physical status of the possible accretion of vacuum energy and vacuum condensates discussed in \cite{Babichev2004,Babichev2005,Cheng2009}  is very desirable. In particular, it was shown in
 \cite{Babichev2004,Babichev2005} that accretion of substance
 with the equation of state of $p=-\rho$ (e.g., Higgs or QCD condensates) decreases a black hole mass, while accretion of the ordinary substance with radiation equation of state increases a black hole mass.

 Investigations of
 these processes in the CUM metric with the applications to an eicheon are waiting. However, one may suggest some scenarios of a galaxy center evolution. Accretion by an eicheon could be more complicated than a traditional black hole. At some stage, eicheon could accrete more ``dark radiation'', increasing its mass, but at some stage, it could accrete more
condensates, decreasing its mass. One could associate this with
the fast processes with a typical time of $\sim 32000$ years. Both
growth of the galaxy's center mass and its lowering produce a
halo. Thus, a galaxy center reminds ``Alice from Wonderland''
\cite{alice}, which takes a bit of a mushroom from one side and
rises then takes a bit from another side and shrinks. These
processes can interlace in a galaxy center.

 To summarize, it is possible to obtain an equation of the state of vacuum polarization, which is some kind of
 ``\ae ther''. It is challenging to find the ``amount'' of \ae ther because it depends on the object's entire history due to the nonlocality of the vacuum state on the curved background. Here we have adjusted this ``amount'' to astrophysical observations. Thus, the obtained final results have, in some sense, a heuristic nature.

\bigskip

{\bf Acknowledgements}

We are grateful to Yoshiaki Sofue for a permission to use the
results of his observation of Milky Way rotational curve in FIG. 4
and FIG 5.

\bibliography{vacpolar}

\end{document}